\documentclass[twocolumn,showpacs,amsmath,amssymb,prl]{revtex4}

\usepackage{graphicx}   
\usepackage{dcolumn}    
\usepackage{bm}         
\newcommand{\rd}{{\rm d}}

\newcommand{\ri}{{\rm i}}

\begin{document}

\title[``Photon''-assisted tunneling]{Analog of photon-assisted tunneling
        in a Bose--Einstein condensate} 

\author{Andr\'e Eckardt}
\author{Tharanga Jinasundera}
\author{Christoph Weiss}
\author{Martin Holthaus}
  
\affiliation{Institut f\"ur Physik, Carl von Ossietzky Universit\"at,
        D-26111 Oldenburg, Germany}
                 
\date{July 21, 2005}

\begin{abstract}
We study many-body tunneling of a small Bose--Einstein condensate in a 
periodically modulated, tilted double-well potential. Periodic modulation of 
the trapping potential leads to an analog of photon-assisted tunneling, with 
distinct signatures of the interparticle interaction visible in the amount of 
particles transferred from one well to the other. In particular, under
experimentally accessible conditions there exist well-developed half-integer 
Shapiro-like resonances.  
\end{abstract}


\maketitle


In a recent experiment, it has become possible to monitor the Josephson-like
oscillation of a sample of about 1000 Bose--Einstein-condensed atoms in an
optical double-well potential, and to observe {\em in situ\/} both the
evolution of the atomic densities, and of the relative phase between the
condensates in both wells~\cite{AlbiezEtAl04}. In the present Letter, we 
propose to extend such experiments by modulating the trapping potential 
periodically in time. As we will demonstrate, such modulated double-well 
condensates exhibit certain parallels to photon-assisted tunneling in 
microwave-driven superconducting Josephson junctions~\cite{BaronePaterno82}, 
although the ``photons'' now tend to fall into the lower Kilohertz regime.

We consider a symmetric double-well trap filled with $N$~Bose particles 
at temperature $T = 0$, as realized approximately in the Heidelberg 
experiment~\cite{AlbiezEtAl04}. Denoting the tunneling splitting between 
the lowest pair of single-particle energy eigenstates by $\hbar\Omega$, 
and adopting the common two-mode
approximation~\cite{MilburnEtAl97,ParkinsWalls98,RaghavanEtAl99b}, the 
system's Hamiltonian takes the idealized form
\begin{equation}
   H_0 = -\frac{\hbar\Omega}{2} \!
   \left( a_1^\dagger a_2 + a_2^\dagger a_1 \right)
   + \hbar\kappa \! \left(
   a_1^\dagger a_1^\dagger a_1 a_1 + a_2^\dagger a_2^\dagger a_2 a_2 \right)
   \; ,
\label{eq:HUN}
\end{equation}
where $a_i^{(\dagger)}$ is the annihilation (creation) operator for a
Boson in the $i$-th well, satisfying the commutation relation
$
   [ a_i, a_j^\dagger ] = \delta_{i,j} \; (i,j = 1,2) .
$
The on-site interaction energy of a single pair of Bosons occupying 
the same well is $2\hbar\kappa$, proportional to the $s$-wave scattering 
length of the particle species. This deceptively simple model~(\ref{eq:HUN}), 
describing an unforced Bosonic Josephson junction, has recently been studied 
in considerable detail~\cite{KalosakasBishop02,KalosakasEtAl03}. The key 
parameter governing its dynamics is the dimensionless ratio $N\kappa/\Omega$.
While the well-understood mean-field 
approximation~\cite{SmerziEtAl97,ZapataEtAl98,RaghavanEtAl99} corresponds
to the limit $N\to\infty$ and $\kappa \to 0$, taken such that the product
$N\kappa$ remains constant, present experiments start to explore the
dynamics beyond the mean-field regime~\cite{AlbiezEtAl04}.
 
We extend this system by introducing a twofold bias: We assume that the 
two wells are tilted such that their bottoms are misaligned in energy by an 
amount $2\hbar\mu_0$, which should be sufficiently large compared to the 
tunnel splitting $\hbar\Omega$, so that the usual Josephson oscillations are 
strongly suppressed. In addition, we propose to modulate this tilted double well 
periodically in time, such that the individual wells are shifted sinusoidally 
up and down, in phase opposition to each other and with angular frequency 
$\omega$, by an amount $\hbar\mu_1$. Such a modulation of an optical double 
well can be achieved by periodically shifting the focus of a blue-detuned 
laser, which creates the barrier between the two wells, with the help of a 
piezo-actuated mirror~\cite{AlbiezEtAl04}. Thus, we are led to the explicitly 
time-dependent many-body Hamiltonian
\begin{equation}
   H(t) = H_0 + \hbar(\mu_0 + \mu_1\sin\omega t)   
   \left( a_2^\dagger a_2 - a_1^\dagger a_1 \right) \; .
\label{eq:HDB}
\end{equation}
As follows from elementary 
estimates~\cite{MilburnEtAl97,ParkinsWalls98,JinasunderaEtAl05}, the 
unperturbed two-mode Hamiltonian~(\ref{eq:HUN}) can be trusted as long as the 
particle number~$N$ does not exceed the ratio of the characteristic length 
scale~$L$ of one of the wells and the scattering length~$a$; for $\mu$m-sized 
traps filled with alkali atoms, this ratio $L/a$ will typically be on the 
order of 1000. However, the ability to reduce atomic scattering lengths by 
means of Feshbach resonances~\cite{InouyeEtAl98} allows one to extend 
the two-mode approximation to much larger samples. Moreover, both Rabi 
frequencies $\mu_0$ and $\mu_1$ and the modulation frequency $\omega$ 
should remain restricted to a few times the tunneling frequency $\Omega$. 
With tunneling times on the order of some 
100 ms~\cite{AlbiezEtAl04,JinasunderaEtAl05}, this leads to modulation
frequencies in the low kHz-regime.    

\begin{figure}
\includegraphics[angle=-90, scale=0.5, width = 7cm]{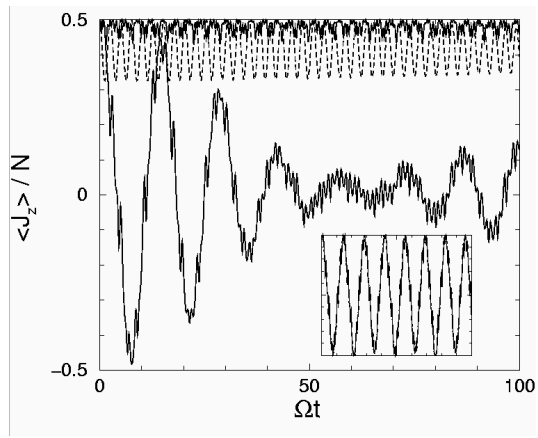}
\caption{Time evolution of the population imbalance $\langle J_z \rangle/N$
        for a Bosonic Josephson junction with $N = 1000$~particles,
        interaction parameter $N\kappa/\Omega = 0.5$, and tilt
        $2\mu_0/\Omega = 3.0$. The dashed line refers to the system
        without periodic forcing; the two full lines are obtained for 
        periodic modulation with one-photon resonant frequency 
        $\omega/\Omega = 3.0$. The strongly oscillating line corresponds 
        to the driving amplitude $2\mu_1/\omega = 1.8$, close to the first 
        maximum of the Bessel function~$J_1$ (inset: solution to the 
        nonlinear Schr\"odinger equation for this situation), while the full 
        line at the top of the figure results for $2\mu_1/\omega = j_{1,1}$, 
        the first zero of $J_1$.} 
\label{F_1}
\end{figure}

In the following, we investigate the dynamics of the model~(\ref{eq:HDB}). 
In our calculations we assume that initially, at time $t = 0$, all $N$~Bose 
particles are prepared in the lower well (that is, in the well tilted downward 
by $-\hbar\mu_0$). When solving the time-dependent Schr\"odinger equation 
numerically, we record the expectation value  
\begin{equation}
   \langle J_z \rangle(t) \equiv \frac{1}{2}\langle \psi(t) | 
   a_1^\dagger a_1 - a_2^\dagger a_2 | \psi(t) \rangle \; ,
\label{eq:PIB}
\end{equation} 
which quantifies the imbalance of the numbers of particles found in the 
individual wells in the course of time, such that 
$\langle J_z \rangle(t_0)/N = +1/2$ 
$(-1/2)$ means that all particles occupy the lower (higher) well at $t_0$. 
We keep the static tilt fixed at three times the tunneling splitting,
$2 \mu_0 = 3 \Omega$. As shown in Fig.~\ref{F_1}, this value allows merely
a small-amplitude oscillation of the population imbalance when there is
no periodic modulation (dashed line).

\begin{figure}
\includegraphics[angle=-90., scale=0.5, width = 7cm]{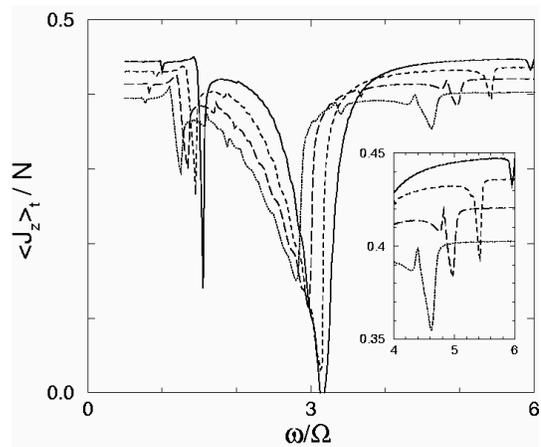}
\caption{Time-averaged population imbalance $\langle J_z \rangle_t/N$
        for the driven Bosonic Josephson junction with $N = 1000$ particles,
        tilt $2\mu_0/\Omega = 3.0$, and scaled driving amplitude
        $2\mu_1/\omega = 0.5$. The interaction parameters are
        $N\kappa/\Omega = 0.1$ (full line), $0.3$ (short dashes),
        $0.5$ (long dashes), and $0.7$ (dots); the averaging interval is
        $\Delta (\Omega t) = 100$. The inset highlights the half-integer 
        resonance at $\omega/2 \approx \omega_0(N\kappa/\Omega)$.} 
\label{F_2}
\end{figure}

In order to condense information about the system's response to bias with 
various parameters in a single viewgraph, we characterize each quantum 
trajectory by its time-averaged imbalance  
\begin{equation}
   \langle J_z \rangle_t \equiv \frac{1}{\Delta t} \int_0^{\Delta t} \!
   \rd t \; \langle J_z \rangle(t) \; ,
\label{eq:TAV}
\end{equation}
employing the averaging interval $\Delta t = 100/\Omega$. Figure~\ref{F_2} 
depicts results of such calculations for a periodic modulation with fixed 
scaled amplitude $2\mu_1/\omega = 0.5$, and frequencies~$\omega$ ranging from 
$0.5\,\Omega$ to $6\,\Omega$. By construction, values of 
$\langle J_z \rangle_t/N$ close to $0.5$ indicate that the condensate remains 
trapped almost entirely in the initially occupied well during $\Delta t$, 
whereas values close to zero signal that it visits both wells about equally. 
From the physics of superconducting Josephson junctions, one expects
photon-assisted tunneling to take place when the energy of an integer number 
of photons matches the static tilt, $n\hbar\omega = 2\hbar\mu_0$, as in the 
case of Shapiro resonances~\cite{BaronePaterno82}. Figure~\ref{F_2} indeed 
reveals a pronounced single-photon resonance at $\omega/\Omega \approx 3.0$, 
clear signatures of two-photon-assisted tunneling at 
$\omega/\Omega \approx 1.5$, and even traces of a three-photon process at 
$\omega/\Omega \approx 1.0$. However, the actual resonance condition does 
not refer directly to the tilt $2\hbar\mu_0$, but rather to the tunneling 
frequency of the condensate without periodic forcing: Within the mean-field 
approximation, the dynamical system~(\ref{eq:HDB}) corresponds to a driven 
nonlinear pendulum~\cite{SmerziEtAl97}; in such a system, a resonance 
can occur when a rationale multiple of the modulation frequency $\omega$ 
matches the oscillation frequency $\omega_0$ of the unforced 
pendulum~\cite{GuckenheimerHolmes90}. This resonance condition 
obviously is effective even when there is no atom-atom interaction. For 
$N\kappa/\Omega = 0$, one has $\omega_0(0)^2 = 4\mu_0^2 + \Omega^2$, 
shifting the one-photon resonance from the Shapiro value 
$\omega/\Omega = 2\mu_0/\Omega = 3$ to $\omega/\Omega \simeq 3.162$. 
Even with weak interaction, $N\kappa/\Omega = 0.1$, this shift remains 
visible in the location of the main resonance peak in Fig.~\ref{F_2}.  

With increasing interaction $N\kappa/\Omega$, the Shapiro-like resonances 
depicted in Fig.~\ref{F_2} shift downwards in frequency. (They shift 
upwards, by different amounts, if initially the higher well is occupied.) 
The mean-field approximation allows one to explain this observation 
quantitatively by expressing the oscillation frequency 
$\omega_0(N\kappa/\Omega)$ of the undriven, interacting system in terms of 
elliptic integrals~\cite{RaghavanEtAl99}. We have checked that the condition 
$n\omega \approx \omega_0(N\kappa/\Omega)$ with $n = 1$, $2$, and $3$ 
reproduces the locations of the one-, two-, and three-photon-like resonances 
quite well.

\begin{figure}
\includegraphics[angle=-90., scale=0.5, width = 7cm]{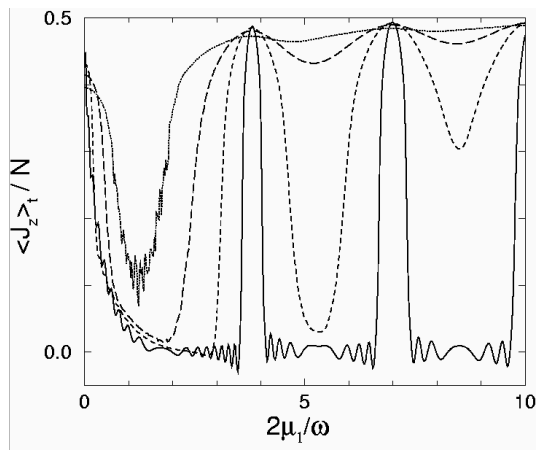}
\caption{Time-averaged population imbalance $\langle J_z \rangle_t/N$
        for the driven Bosonic Josephson junction with tilt 
        $2\mu_0/\Omega = 3.0$ and scaled driving amplitude
        $2\mu_1/\omega = 1.8$. The full line is obtained from the
        nonlinear Schr\"odinger equation for $N\kappa/\Omega = 0.1$;
        the long-dashed line refers to a non-interacting system. 
        Also shown is the result for a system with only $N = 2$ particles,
        keeping $N\kappa/\Omega = 0.1$ fixed (short dashes; displaced
        downwards by 0.1 for clarity).}  
\label{F_3}
\end{figure}

Figure~\ref{F_2} also reveals a well-developed {\em half-integer} resonance 
at $\omega/2 \approx \omega_0(N\kappa/\Omega)$, corresponding to $n = 1/2$. 
Such subharmonic resonances can again be understood with the help of the 
nonlinear Schr\"odinger equation, which describes the dynamics of the driven 
condensate within the mean-field approximation: The nonlinearity effectively 
generates both harmonics and subharmonics of the driving frequency. For a 
matter-of-principle demonstration, Fig.~\ref{F_3} shows the time-averaged 
imbalance provided by the nonlinear Schr\"odinger equation for stronger 
forcing with scaled amplitude $2\mu_1/\omega = 1.8$ and rather weak 
interaction $N\kappa/\Omega = 0.1$, together with the corresponding data for 
a non-interacting system subjected to the same forcing. While the mean-field 
result exhibits half-integer resonances with $n = 3/2$ and $n = 5/2$, these 
resonances vanish when $N\kappa/\Omega = 0$. On the other hand, the graph of 
the imbalance~(\ref{eq:TAV}) provided by the $N$-particle Schr\"odinger 
equation for $N = 1000$ (not shown) is practically indistinguishable from that 
of the mean-field data displayed in Fig.~\ref{F_3}, and even a two-particle 
system yields a pronounced peak for $n=3/2$ when $N\kappa/\Omega$ is adjusted 
to $0.1$. In general, while the time-dependent $N$-particle 
imbalance~(\ref{eq:PIB}) follows the mean-field data only in the limit 
$N \to \infty$ (with $\kappa$ adjusted such that $N\kappa$ stays constant), 
but soon deviates from its mean-field counterpart when $N$ is of the order of 
$1000$ (cf.\ Fig.~\ref{F_1}), it is remarkable that its average~(\ref{eq:TAV}) 
comes fairly close to the mean-field approximation even when $N$ is 
{\em much\/} smaller.

A powerful theoretical tool for studying the many-body dynamics of the 
driven Bosonic Josephson junction is provided by quantum Floquet theory:
Since the Hamiltonian~(\ref{eq:HDB}) is periodic in time, $H(t) = H(t+T)$
with period $T= 2\pi/\omega$, there exists a complete set of solutions
to the time-dependent Schr\"odinger equation of the form
$
   |\psi_m(t)\rangle = |u_m(t)\rangle 
        \exp\!\left(-\ri \varepsilon_m t/\hbar \right) ,
$       
with the Floquet functions $|u_m(t)\rangle = |u_m(t+T)\rangle$ sharing 
the $T$-periodicity of the Hamiltonian~\cite{Shirley65}. The quantities 
$\varepsilon_m$, which serve to describe the time evolution in a similar 
manner as energy eigenvalues do in the case of energy eigenstates, are called 
quasienergies. The existence of these time-dependent many-body states, which 
incorporate both the interparticle interactions and the external forcing in
a non-perturbative manner, is not tied to the particular model 
Hamiltonian~(\ref{eq:HDB}), but stems solely from the temporal periodicity  
of the forcing, exactly as the existence of Bloch states in solid state 
physics is solely due to the spatial periodicity of a crystalline 
lattice~\cite{AshcroftMermin76}.

While the exact calculation of Floquet states and their quasienergies
in general has to ressort to numerical means~\cite{JinasunderaEtAl05},
the particular case of an integer resonance, $n \omega = 2\mu_0$,
allows for an instructive analytical approximation. Provided the static 
tilt is sufficiently large to substantially reduce the unperturbed
Josephson oscillations, and assuming low resonance order~$n$, so that
$\omega/\Omega$ remains large compared to $N\kappa/\Omega$, the 
{\em quasi\/}energies of the {\em driven\/} junction~(\ref{eq:HDB}) are
approximately given by the {\em energy\/} eigenvalues of the {\em undriven\/}
junction~(\ref{eq:HUN}), but with an effective tunneling frequency     
\begin{equation}
   \Omega_{\rm eff} = \Omega J_n\!\left(\frac{2\mu_1}{\omega}\right) \; ,
\label{eq:REN}
\end{equation}
where $J_n(z)$ is an ordinary Bessel function of order~$n$. This 
renormalization of the tunneling frequency under resonant driving
resembles the renormalization of atomic $g$-factors by radiofrequency
fields~\cite{HarocheEtAl70}, or the coherent destruction of
single-particle tunneling~\cite{GrifoniHanggi98}.  

To demonstrate that this concept is meaningful under experimentally 
accessible conditions, we haved displayed in Fig.~\ref{F_1} the 
evolution of the population imbalance~(\ref{eq:PIB}) under the influence
of a strong modulation: The lower, strongly oscillating full line results 
from one-photon resonant driving with scaled amplitude $2\mu_1/\omega = 1.8$, 
which is close to the first maximum of $J_1$. Correspondingly, one observes 
photon-assisted tunneling with almost complete population exchange during 
the initial stage, leveling down to smaller-amplitude oscillations which 
average to zero. As remarked above, the solution to the nonlinear Schr\"odinger
equation for the same situation (inset) deviates strongly from the $N$-particle
solution after only a few periods. The upper full line in Fig.~\ref{F_1} 
results from one-photon resonant driving with $2\mu_1/\omega \simeq 3.832$, 
the first zero of $J_1$: Since the effective tunneling frequency~(\ref{eq:REN})
vanishes then, the particles remain caught almost perfectly in the initially 
occupied well.

\begin{figure}
\includegraphics[angle=-90., scale=0.5, width = 7cm]{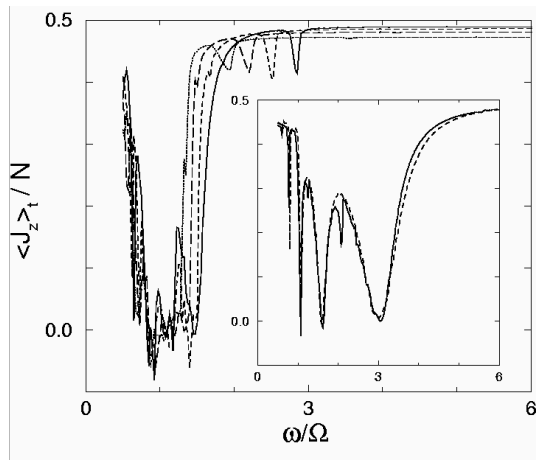}
\caption{Time-averaged population imbalance $\langle J_z \rangle_t/N$
        for the driven Bosonic Josephson junction with $N = 100$ particles,
        tilt $2\mu_0/\Omega = 3.0$, and one-photon resonant driving
        frequency $\omega/\Omega = 3.0$. The interaction parameters are 
        as in Fig.~\ref{F_2}.}
\label{F_4}
\end{figure}

The renormalization~(\ref{eq:REN}) prompts at an interplay between
photon-assisted tunneling and an effect known as self-trapping. It has been 
established that when the absolute value of the interaction parameter 
$|N\kappa/\Omega|$ exceeds a certain critical value, which equals $1$ for 
maximum initial population imbalance, the nonlinear Josephson oscillations 
occuring in the undriven junction become 
suppressed~\cite{SmerziEtAl97,RaghavanEtAl99}; 
this self-trapping phenomenon has by now been observed in the pioneering 
experiment~\cite{AlbiezEtAl04}. Therefore, employing driving frequencies
sufficiently high for the  modulation-induced renormalization~(\ref{eq:REN}) 
to hold, and starting from an initial state with all particles in one well, 
we expect self-trapping to occur in a resonantly driven Bosonic Josephson 
junction when 
\begin{equation}
   \left| \frac{N\kappa}{\Omega \, J_n(2\mu_1/\omega)}\right| > 1 \; . 
\label{eq:CON}
\end{equation} 
This expectation is confirmed in Fig.~\ref{F_4}, where we have plotted
time-averaged population imbalances as functions of the scaled driving
amplitude, again for the example of a one-photon resonance. The number 
of particles has been reduced to $N = 100$ here: As far as the 
averages~(\ref{eq:TAV}) are concerned, this yields practically the same
results as $N = 1000$, and even as the nonlinear Schr\"odinger equation,
despite the differences illustrated in Fig.~\ref{F_1}. For very low 
driving amplitudes, $2\mu_1/\omega < 0.1$, the forcing is hardly effective, 
so that the static tilt keeps the particles in the initially occupied well. 
With increasing amplitude, photon-assisted tunneling sets in. For 
$2\mu_1/\omega$ larger than about $1.5$, the four curves drawn in 
Fig.~\ref{F_4} behave rather differently: For low interaction parameter 
$N\kappa/\Omega = 0.1$ (full line), the trapping condition~(\ref{eq:CON}) is 
effective only in comparatively small neighborhoods of the zeros 
$j_{1,1} \simeq 3.832$, $j_{1,2} \simeq 7.016$, and $j_{1,3} \simeq 10.173$ 
of $J_1$. When $N\kappa/\Omega$ is increased to $0.3$ (short dashes), 
these neighborhoods become wider; in the entire interval between $j_{1,2}$ 
and $j_{1,3}$ the condition~(\ref{eq:CON}) is met, so that the imbalance does 
not approach zero. For $N \kappa/\Omega = 0.5$ (long dashes) self-trapping 
occurs already between the first and second zero, while $N \kappa/\Omega = 0.7$
(dots) does not admit full population exchange even for scaled amplitudes 
below $j_{1,1}$, in agreement with the inequality~(\ref{eq:CON}). Thus, 
the interactions among the Bose particles exert a strong influence on 
photon-assisted tunneling, leading to a distinctly non-monotonous dependence 
of the population transfer on the driving amplitude, and allowing substantial 
transfer only in limited subsets of the parameter space.

In conclusion, we have suggested to extend the on-going experimental 
investigation of condensate tunneling in optical double 
wells~\cite{AlbiezEtAl04} by a time-periodic modulation of the trapping 
potential, thereby opening up parallels to the ac Josephson effect. 
Many-body Floquet theory leads to the prediction of the amplitude-dependent 
breathing~(\ref{eq:REN}) of the well-to-well coupling under strong resonant 
driving. However, driven Bosonic Josephson junctions offer further 
peculiarities of their own, such as interaction-induced resonance shifts, 
and clearly developed half-integer Shapiro-like resonances, which should be 
verifiable under presently accessible laboratory conditions. Subharmonic 
Shapiro steps in microwave-driven superconducting point contacts have recently 
been ascribed to multiple Andreev reflections~\cite{CuevasEtAl02}. The fact 
that related phenomena occur in modulated optical Bosonic junctions, where all
of the  parameters $N$, $\Omega$, $\kappa$, $\mu_0$, $\mu_1$, and $\omega$ can 
be tuned separately, opens up far-reaching new avenues of investigation. 
        
This work was supported by the Deutsche Forschungsgemeinschaft through 
the Priority Programme SPP~1116, {\em Wechselwirkung in ultrakalten Atom- 
und Molek\"ulgasen\/}. A.E.\ acknowledges a fellowship from the
Studienstiftung des deutschen Volkes.

\end{document}